\documentclass[conference]{IEEEtran}

\usepackage{graphicx}  

\usepackage{subfigure} 
\begin{document}

\title{Anisotropy Studies with the Pierre Auger Observatory}

\author{\authorblockN{E. M. Santos\authorrefmark{1},
for the Pierre Auger Collaboration}
\\
\authorblockA{\authorrefmark{1}Laboratoire de Physique Nucl\'eaire et de Hautes \'Energies, 
 T33 RdC, 4 place Jussieu, 75252 Paris Cedex 05, France \\ Email: emoura@lpnhe.in2p3.fr}}


\maketitle

\begin{abstract}
An anisotropy signal for the arrival directions of ultra-high energy cosmic rays (UHECR) of more 
than 99\% confidence level was established using data collected by the Pierre Auger 
Observatory. Cosmic rays with energy above $\sim 6 \times 10^{19}$ eV show a correlation with 
the positions of extragalactic nearby active galactic nuclei (AGN), being maximum for 
sources at less than $\sim$100 Mpc and angular separation of a few degrees. The evolution 
of the correlation signal with the energy shows that the departure from anisotropy coincides 
with the flux suppression observed in the spectrum, being therefore consistent with the hypothesis 
that the correlated events have their origin in extragalactic sources close enough to 
avoid significant interaction with the cosmic microwave background (the Greisen-Zatsepin-Kuz'min 
effect). Even though the observed signal cannot unambiguously identify AGNs as the production sites 
of UHECRs, the potential sources have to be distributed in a similar way. A number of additional 
statistical tests were performed in order to further understand the nature of the correlation signal.
\end{abstract}


%
\IEEEpeerreviewmaketitle

\section{Introduction}

The origin of the highest energy cosmic rays (above $10^{18}$ eV), as far as their mass composition 
and production mechanisms are concerned, is still undetermined. However, the recent Auger results, 
which include limits on the diffuse flux of neutrinos \cite{Abraham:2007rj}, bounds on the flux of 
photons above $10^{19}$ eV obtained with the Surface Detector (SD) and Fluorescence Detector (FD) 
together \cite{Abraham:2006ar} and with the SD only \cite{Aglietta:2007yx}, have put stringent 
bounds on top-down models, favoring, therefore, the hypothesis of production by acceleration process 
in powerful astrophysical sources \cite{Semikoz:2007wj}.

Anisotropy studies play an important role in the identification of the sources and can even 
provide valuable information on the chemical composition. Detectors like the large Auger surface 
array are sensitive to showers with energies around above $10^{17}$ eV, but the detector reaches 
100\% efficiency only above $3\times 10^{18}$ eV. Given the steep falling spectrum, even taking 
into account the rapidly falling efficiency of the detector below this threshold, the Auger Surface 
Detector sample is dominated by below-saturation showers. In this energy region, despite the higher 
statistics, the myriad of systematic effects associated to the detector exposure makes large 
scale anisotropy analysis a very challenging task \cite{Santos:2008rv}. Weather induced effects, 
like seasonal and diurnal trigger rate modulations \cite{Bleve:2007mb}, geometric effects 
associated to the particular arrangement of the SD array and the influence of the local geomagnetic 
field on the air shower development are some of the examples of systematic effects which one has to 
deal with in order to extract reliable information on anisotropies over extended regions of the 
sky. For energies above $10^{18}$ eV, the Pierre Auger Collaboration has already published some 
results on one-dimensional analyzes, that is, right-ascension first harmonic analysis, Fourier 
development in modified sidereal time \cite{Billoir:2007nu} and the East-West method 
\cite{Armengaud:2007hc}. The galactic centre is a natural candidate for searches of anisotropy, 
since there are a number of evidences that it harbors, as essentially every galaxy in the universe, 
a massive black hole, which would be associated to the strong radio emissions from Sagittarius A*, 
and the $\gamma$-rays detected by H.E.S.S. \cite{Aharonian:2004wa}. However, above $10^{17}$ eV, 
the Auger results in the galactic centre region show no signs of anisotropy \cite{Aglietta:2006ur}, 
despite claims in the past of large excesses in this region of the sky 
\cite{Hayashida:1998qb,Bellido:2000tr}.

At the very edge of the cosmic ray spectrum measured so far, the systematic effects are no longer 
so dominant, however, now the extremely low statistics is the challenge to be faced. On the other 
hand, it has been long since people realized that at such extreme energies, charged particles 
magnetic rigidity, both in galactic and extra-galactic fields, should be large enough to avoid a 
complete isotropisation of their arrival directions. In fact, the Auger Collaboration has recently 
demonstrated that its highest energy events, that is, above 60 EeV (1 EeV = $10^{18}$ eV), are 
anisotropic \cite{Cronin:2007zz,Abraham:2007si}. The result has been established by means of a 
correlation found between the arrival directions of the showers and a catalog of AGNs. The result 
has demonstrated the feasibility of charged particle astronomy. This conference paper contains a 
discussion of the results presented in the two papers mentioned above and some remarks on their 
implications to the cosmic ray physics as well as on issues they have raised after publication. 
\begin{figure}[h]
\centering
\includegraphics[width=3.4in]{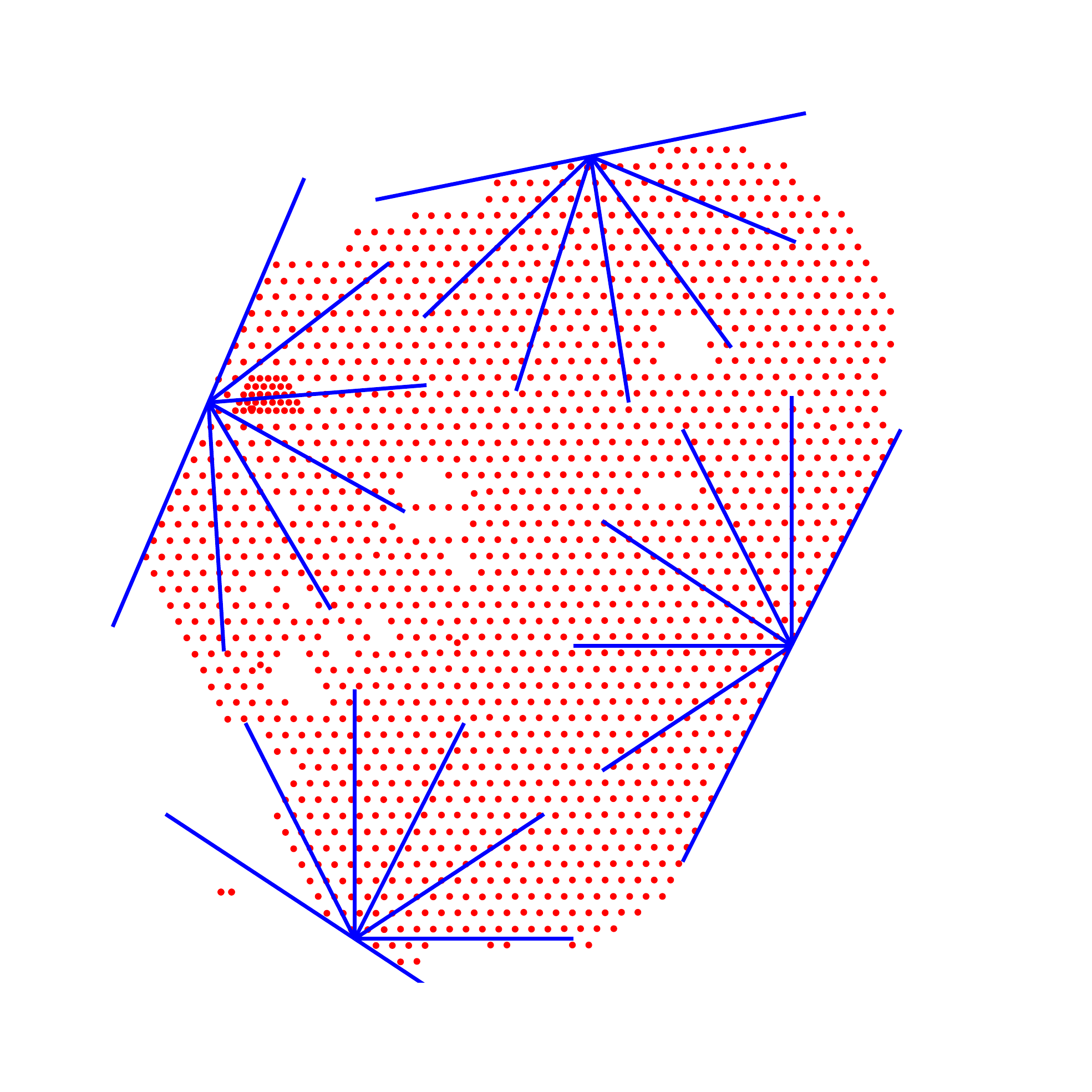}
\caption{Surface detector array status as of 27 October 2008, showing a total of 1638 stations 
deployed and taking data. The field of view of the four fluorescence eyes (with 6 telescopes each) 
are represented by the blue lines (each one is 20 km long). Some infills, tanks deployed at half 
the distance of the regular array from each other in order to lower the observatory energy threshold 
can also be seen close to one of the eyes.}
\label{array_fig}
\end{figure}

\section{Current Array Status and the Data Set Used}
The Pierre Auger Observatory has now reached completion with all of its 1600 surface detector 
stations deployed on ground and the FD with all 24 telescopes taking data 
together since February 2007. The surface detector has been taking data stably since January 2004. 
The data presented in this conference contribution corresponds to showers detected by the SD in 
the period 1 January 
2004 up to 31 August 2007, amounting for a total integrated exposure of 9000 km$^2$ sr yr. In this 
period, the surface detector has continuously grown in size from only 154 to 1388 stations.

In order to avoid border effects, which tend to be important during the detector growing process, 
when the array has a non negligible number of holes, we consider only events where the station 
with the highest signal is surrounded by at least 5 other active tanks, that is, stations actually 
taking data when the shower reaches the ground. To further constrain the impact of dead stations 
in the event reconstruction, the estimated shower core position is required to fall inside a 
triangle of active stations. 
\begin{figure*}
\centerline{\subfigure[]{\includegraphics[width=3.0in]{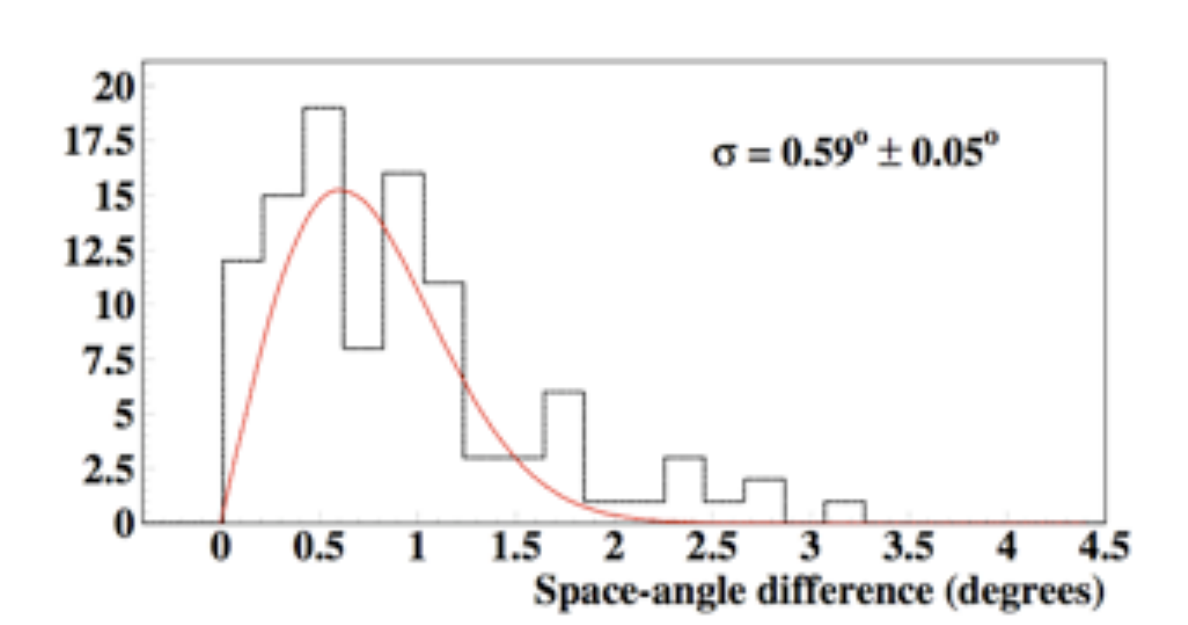}
\label{fig_ardoublets}}
\hfil
\subfigure[]{\includegraphics[width=3.0in]{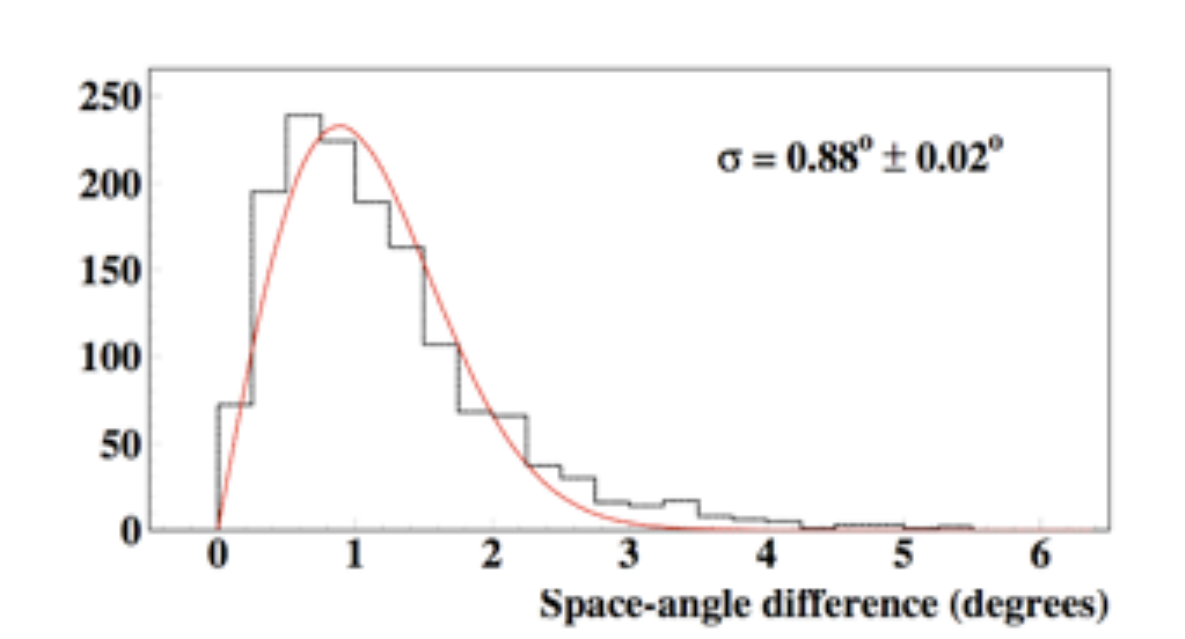}
\label{fig_arhybrids}}}
\caption{(a) Distribution of the space angle between reconstructions of the same shower performed 
with independent sets of tanks in a sub-array of doublets. Defining the AR as the angle containing 
68\% of all reconstructed showers coming from a point source, AR=1.5$\sigma$. (b) Distribution of 
the space angle between the reconstruction done using only the surface detector and the surface and 
fluorescence detectors together (the hybrids). Notice that the intrinsic hybrid resolution contributes 
to the total dispersion observed in this plot.}
\end{figure*}

Obviously, for anisotropy studies, the accuracy with which one reconstructs the shower direction 
plays a special role. The reconstruction of the event direction is done by fitting the arrival 
times predicted by a 
certain shower front model (either a plane or a curved form) propagating at the speed of light 
to the measured arrival times in the tanks triggered by a shower. The accuracy achieved in 
such a reconstruction is limited essentially by the clock resolution of the station and the 
intrinsic fluctuations of the arrival time of the first particle \cite{Bonifazi:2007ck}. The 
Auger SD angular resolution (AR) above 10$^{19}$ eV, is estimated to be better than 1$^{\circ}$, 
as defined by the angular window which would contain 68\% of all reconstructed showers produced by 
primaries coming from a point source \cite{Ave:2007wf}. Figures \ref{fig_ardoublets} and 
\ref{fig_arhybrids} show two different estimations of the AR in Auger, using 
a sub-array of doublets\footnote{Doublets are pairs of tanks deployed very 
close to each other (11 m), so that they sample essentially the same part of the shower.}, 
through which is possible to reconstruct the same shower with two independent sets of tanks 
and using the hybrid events, by comparing the SD and SD+FD reconstructions. 

By profiting from the unique hybrid nature of the Auger detector, the so called hybrid events, the 
ones detected simultaneously by the SD and the FD, are used to inter-calibrate these two detectors, 
providing an energy estimation which is almost independent of Monte Carlo simulations. Firstly, we 
have identified a SD observable which could be 
used to estimate the shower energy, and this was taken be the estimated signal 1000 m ($S(1000)$) 
from the reconstructed shower core. Secondly, as such a signal is attenuated by the atmosphere, a 
correction is applied using the so called Constant Intensity Cut (CIC) \cite{Hersil:1961zz}. After 
the correction, $S(1000)$ is normalized to a reference angle 
(38 degrees) and the final observable ($S_{38^{\circ}}$) can then be correlated to the calorimetric 
energy measurement performed by the FD \cite{Abraham:2008ru}. Once such a calibration curve is built 
for the hybrid events, it can be used for the whole high statistics sample measured by the SD.

%

\section{Establishing the Anisotropy of the Highest Energy Sample}

\subsection{The source catalog and the scan method}
The Veron-Cetty (V-C) catalog \cite{VC} is a compilation of different surveys and its 12th 
edition contains 85221 quasars, 1122 BLLacs and 21737 AGNs (of which 694 are less 
than approximately 100 Mpc from us). By considering only its closest sources 
($z\le 0.024$), the probability for chance correlations with sources from the catalog, 
already taken into account the Auger exposure, is as high as approximately 70\% for 
angular windows of 7$^{\circ}$. In other words, it is still possible to look for 
anisotropies. As farther sources are aggregated, of course, chance correlation becomes 
essentially 100\%, making searches for deviations from isotropy impossible.

Anisotropy signals in the cosmic ray arrival directions might be identified with the 
help of an intermediate source catalog through which a correlation signal might be 
established using an appropriate set of parameters such as: the typical angular scale 
of the correlation signal, the maximum redshift of the candidate sources and the 
minimum energy of the events. A scan over these parameters can then 
be performed looking for the typical values leading to a minimum of the probability 
for such a data-catalog correlation has been produced by chance. Given $N$ events 
above a certain energy threshold $E_{min}$, and a set of sources (limited by a maximum 
redshift $z_{max}$), the probability that at least $k$ of them are at less than a certain 
angular distance $\gamma$ from one of the catalog's sources, just by chance, is given 
by the cumulative binomial probability
\begin{equation}
P = \sum_{j=k}^{N}\left(
\begin{array}{c}
N \\
j
\end{array}
\right) p^{j}(1-p)^{N-j},
\end{equation}
where $p(\gamma, z_{max})$ is the probability for an event, drawn uniformly on the sky 
(but taking into account the detector exposure), to be at an angular distance less than 
$\gamma$ from at least one of the sources with $z\le z_{max}$.

\subsection{The exploratory scan}
A potential anisotropy signal in a subset of the Auger SD data, collected between 1 January 
2004 and 27 May 2006, was identified through a scan over the 3 parameters mentioned above. 
Angular windows from 1$^{\circ}$ to 6$^{\circ}$, V-C sources up to $z=0.024$ 
and showers with energies above 40 EeV were considered. The minimum and maximum values 
for the windows are limited by the detector angular resolution and the constraint of 
moderate chance correlation probabilities with sources up to a certain maximum redshift, 
respectively. The soil energy was chosen in order to cope with the fact that typical magnitudes 
for the galactic and extra-galactic magnetic fields, will imprint deviations on low energy 
primaries, even protons, larger (in average) than the angular interval considered. The 
minimizing parameters for this exploratory scan were: $\gamma=3.1^{\circ}$, $z_{max}=0.018$ 
and $E_{min}=56$ EeV. For such a configuration, 12 out of 15 events were correlated with 
V-C sources when only 3.2 were expected by chance (21\% of chance correlation probability). 

\subsection{The signal confirmation}
The Pierre Auger Collaboration has an internal policy which states that any potential 
anisotropy signal should be tested on an independent data set \cite{Clay:2003pv}. Therefore, 
a prescription was established in order to confirm or to reject the potential signal. 
Given the very low statistics at the energies considered, the Collaboration decided to 
proceed through a sequential analysis with a predefined stopping rule, since such tests 
can give an answer at earlier stages in the sequence. We have chosen {\it isotropy} as our 
{\it null hypothesis}, adopting a $\alpha=1$\% Type I error, that is, the allowed probability 
for a false claim of anisotropy, and a Type II error of $\beta=5$\%, that is, the probability 
for an incorrect claim that the data was isotropic. The error $\beta$ determines the power 
of the test ($1-\beta$), that is, $95$\%. By moving the bounds of the exploratory scan, 
the correlation power could drop from 80\% (12/15) to values as low as 60\% (8/14). In 
order to keep the test power above the 95\% threshold, even in the case of such a lower 
correlation power, the total length of the test was set to 34 additional detected events 
above 40 EeV (the predefined stopping rule).

This prescription was fulfilled almost a year later on 25 May 2007, with the detection 
of 8 new events above 40 EeV, of which 6 were at less than 3.1$^{\circ}$ from AGNs up to 75 Mpc 
from us. We could then claim that the hypothesis of isotropy for the highest energy Auger events 
($E>40$ EeV) was rejected with at least 99\% CL.

\begin{figure}
\centering
\includegraphics[width=3.4in]{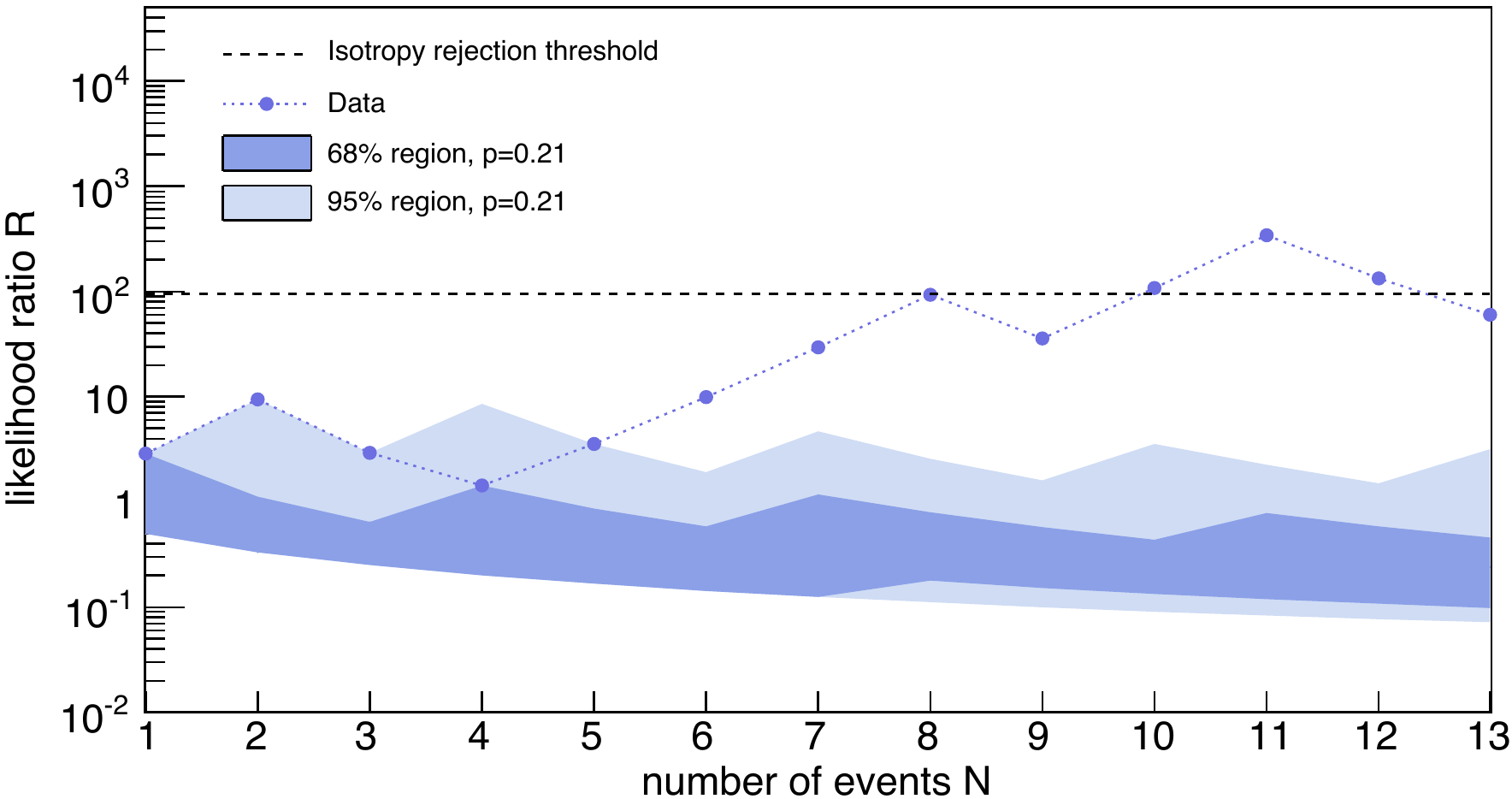}
\caption{Likelihood ratio $R$ (see eq. \ref{lratio}) evolution for every new event detected 
since the start of the internal Auger prescription. The threshold for isotropy 
rejection is represented by the dashed horizontal line. The 68\% and 95\% CL regions from 
isotropic expectations are also shown. Isotropy rejection was achieved after 10 detected 
events.}
\label{fig_lratio}
\end{figure}

Hypothesis tests done via a classical sequential analysis have to deal with the fact that a penalty 
factor should be included at each intermediate step where the hypothesis has been tested, 
simply due to the fact that by performing the test several times, the probability for 
fulfilling it by chance is enhanced. Such a penalty factor will therefore depend on the total 
length of the test, introducing an undesirable dependence of the corresponding probability 
being monitored on data actually not taken. These classical tests drawbacks can be avoided 
by the use of a likelihood ratio test \cite{Wald:1945}. In figure \ref{fig_lratio} one can see the 
evolution of the likelihood ratio \footnote{The numerator is the probability to get a particular 
data set in case the anisotropy hypothesis is true and the denominator the corresponding probability 
when the isotropy hypothesis is true. Notice that since we do not know the true correlation power 
$p_{1}$ in case of anisotropy, we have integrated over this variable.} \cite{BenZvi:2007gr}
\begin{equation}
R = \frac{\int_{p}^{1}{p_{1}^{k}(1-p_{1})^{N-k}dp_{1}}}{p_{k}(1-p)^{N-k+1}},
\label{lratio}
\end{equation}
as a function of time and the rejection of the isotropy hypothesis after the collection of 
10 events out of which 7 were correlated with AGNs for the prescription parameters.

%

\begin{figure*}
\begin{center}
\includegraphics[width=3.4in]{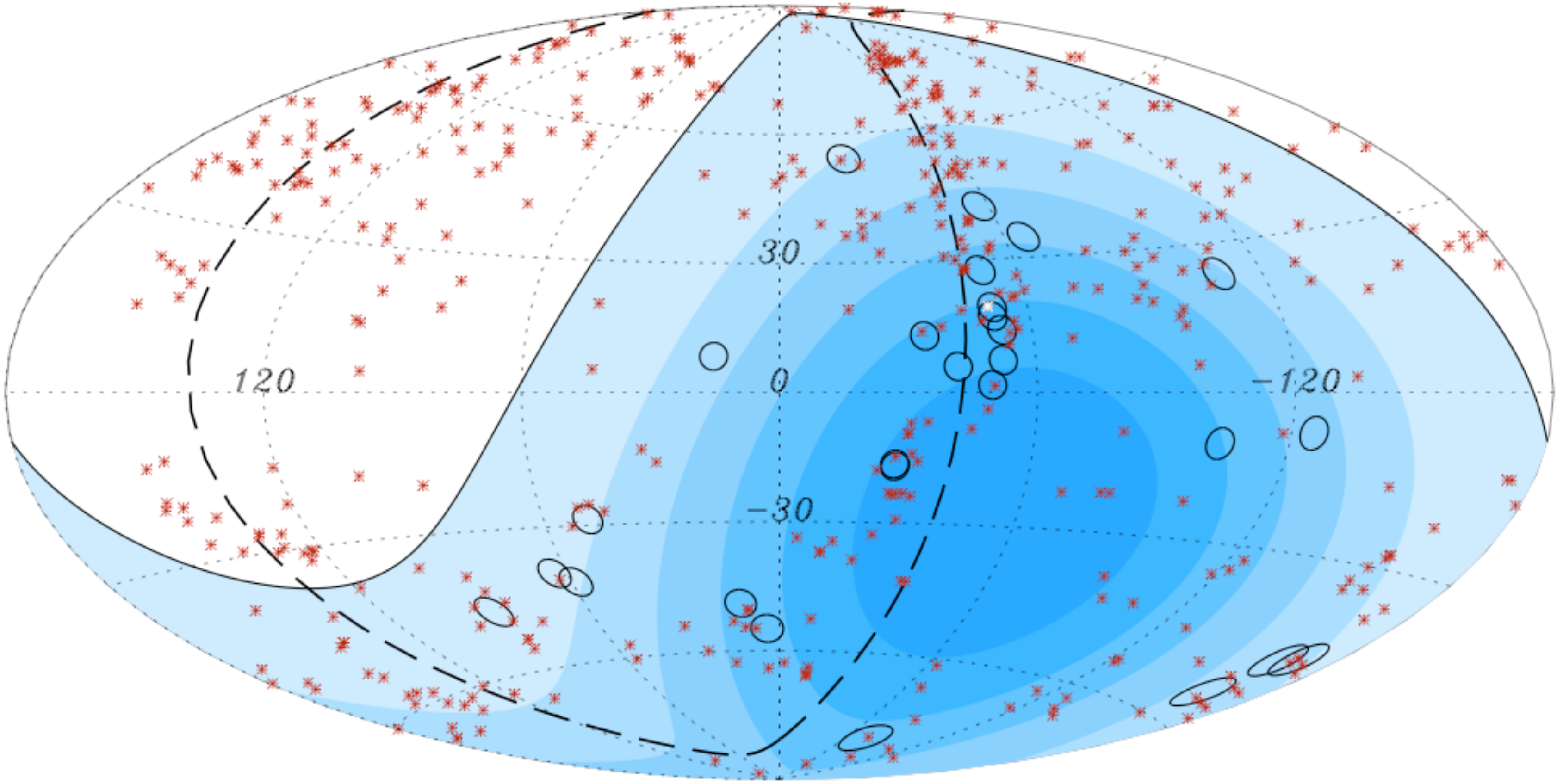}
\caption{Sky map in galactic coordinates showing the directions of 27 Auger 
highest energy events ($E>57$ EeV) with angular windows of 3.2 $^{\circ}$ 
around them, as well as the positions of the 442 AGNs in the V-C catalog 
with redshift $z\le 0.017$ ($\sim 71$ Mpc). The color code indicates the observatory 
relative exposure over the sky. Centaurus A is marked in white.}
\label{fig_skymap}
\end{center}
\end{figure*}

\subsection{Combining the full data set}
Once the anisotropy signal has been confirmed with the independent data set, all 
the data collected since 1 January 2004 was submitted to the same scan, using as 
well a new calibration available at the time. The minimum was achieved for the 27 
events above 57 EeV, of which 20 were at less than 3.2$^{\circ}$ from AGNs of the 
V-C catalog up to approximately 71 Mpc ($z_{max}=0.017$). Only 5.6 events are expected 
to correlate by chance under such conditions, and the probability for that to have 
happened by chance (already penalizing for the scan procedure) is approximately 
$10^{-5}$.

Figure \ref{fig_skymap} contains a sky map in galactic coordinates with the 27 events 
above 57 EeV for the full data set with circles of 3.2$^{\circ}$ around each of them, 
as well as the positions of the 442 AGNs with $z\le 0.017$. The color code indicates 
the Auger relative exposure in each direction of the sky. A clear property of the 
events distribution is the concentration close to the supergalactic plane (dashed line), 
following the local anisotropic matter distribution in our neighborhood. Notice as well 
the two events at less than 3$^{\circ}$ from Centaurus A, one of the closest AGNs.

\section{Signal Properties}

A closer look at the signal behavior in the vicinity of the minimum in the parameter 
space, shows the presence of additional secondary minima, indicating that the minimizing 
parameters shown above should not be taken at their face values, but only as indicative 
of the true correlation scales. Another distinctive property of the signal evolution 
as a function of energy is the sharp transition to the global minimum, suggesting that 
something abrupt happens around $E_{min}$ and have the effect of enhancing the anisotropy 
signal. 

%
\begin{figure*}
\centerline{\subfigure[]{\includegraphics[width=2.9in]{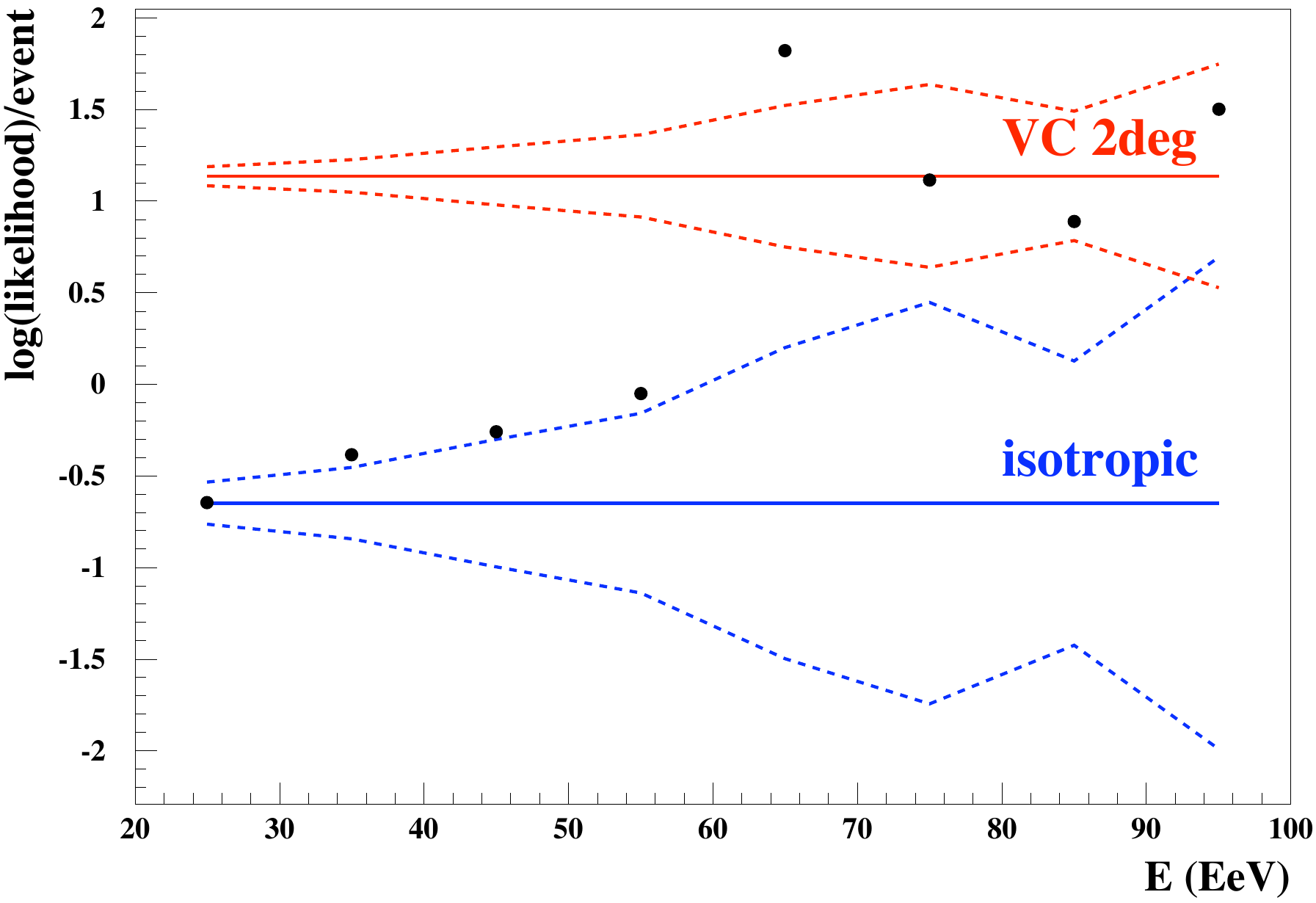}
\label{fig_loglikeli}}
\hfil
\subfigure[]{\includegraphics[width=2.9in]{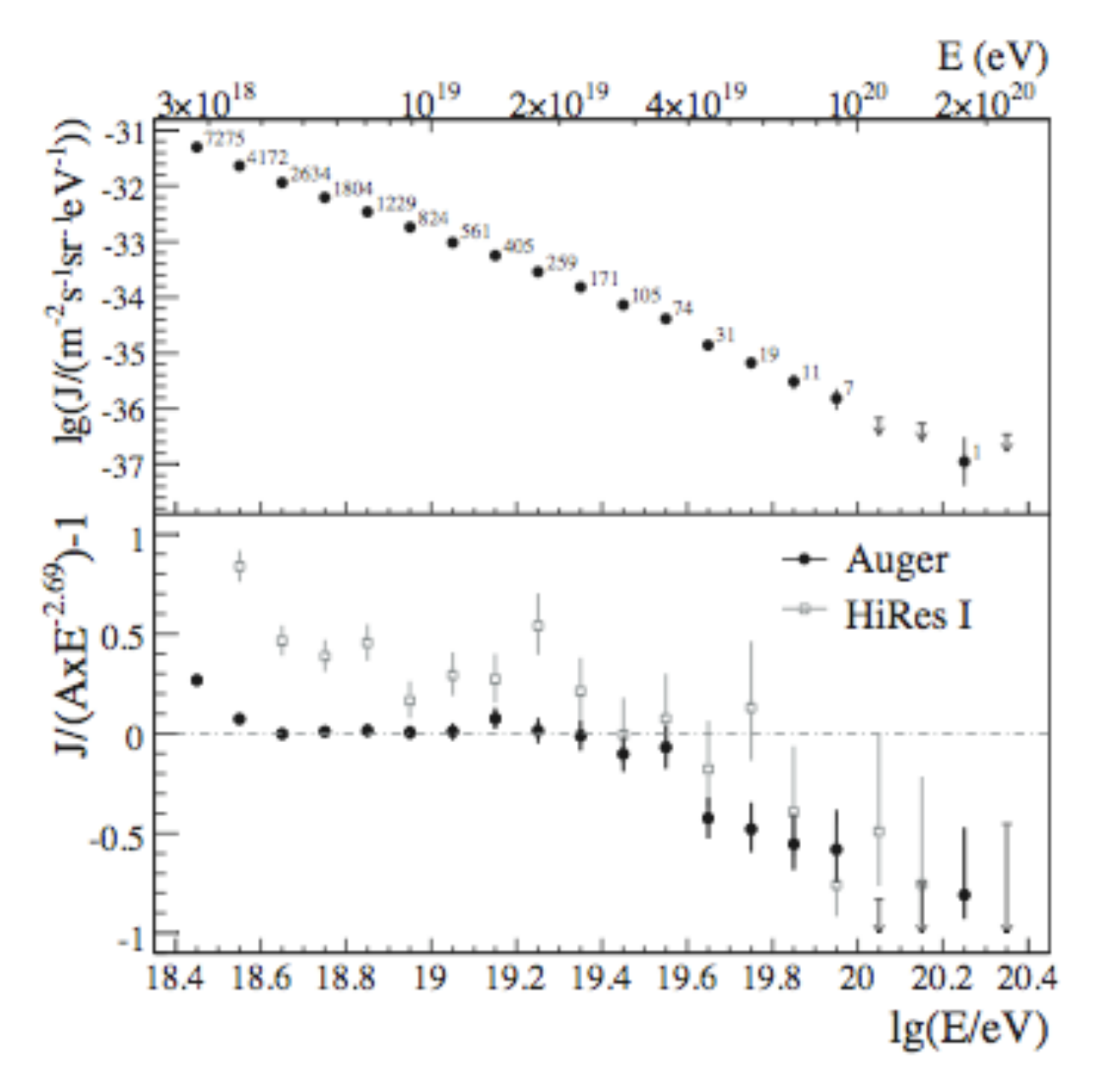}
\label{fig_spectrum}}}
\caption{(a) Average log-likelihood per event (dots) for the Auger events as function 
of the energy threshold in EeV compared to the expectations from samples drawn from the 
V-C catalog smoothed at a 2$^{\circ}$ scale (dashed red line) and from isotropic samples 
(dashed blue line). The solid lines represent 1$\sigma$ bounds. (b) top: Differential 
flux as a function of energy and its 
statistical uncertainty. The number of detected showers for each data point is shown on 
top of each point; bottom: Fractional difference between the Auger cosmic ray SD 
spectrum and an $E^{-2.69}$ power law. Notice the coincidence between the onset of 
anisotropy for the events above $\sim$ 60 EeV in the plot on the left-hand side and the 
energy where the flux drops by a factor 2 from what would be expected from a power 
law on the right-hand side. The HiRes data is also shown for comparison.}
\label{fig_anispec}
\end{figure*}

Figure \ref{fig_loglikeli} shows the average log-likelihood per event LL as a 
function of the energy threshold, defined as
\begin{equation}
LL = \frac{1}{N}\sum_{k=1}^{N} \log(\rho_{k}),
\end{equation}
where the sum runs over all the events above the energy threshold and $\rho_{k}$ 
is the column density in the direction of the $k$-event of a map built by smoothing 
out the V-C catalog with Gaussians of 2$^{\circ}$ around each of its sources. The plot 
shows also the 1$\sigma$ expectation bounds of the log-likelihood for sets of events 
(consistent with the statistics in the data for the corresponding energy threshold) 
drawn from the V-C catalog itself (red lines) and from an isotropic distribution 
(blue lines). One can clearly see the abrupt transition from isotropy to V-C-like 
anisotropy as the energy reaches 60 EeV from below. Such a sharp transition 
is accompanied by a 50\% drop in the flux, from what would be expected from 
a simple extrapolation of the measured flux just below the cutoff, in the form of 
a $E^{-2.69}$ power law as can be seen in figure \ref{fig_spectrum}. This striking 
coincidence is an additional evidence for the existence of the GZK effect, in which 
the sudden horizon shrinking, as the energy threshold for pion photo-production is 
reached, has the immediate consequence of revealing the anisotropies of our clumpy 
neighborhood. If sources were simply running out of power at energies around 60 EeV, 
such a sudden appearance of anisotropy would not be expected.

\begin{figure}
\centering
\includegraphics[width=3.2in]{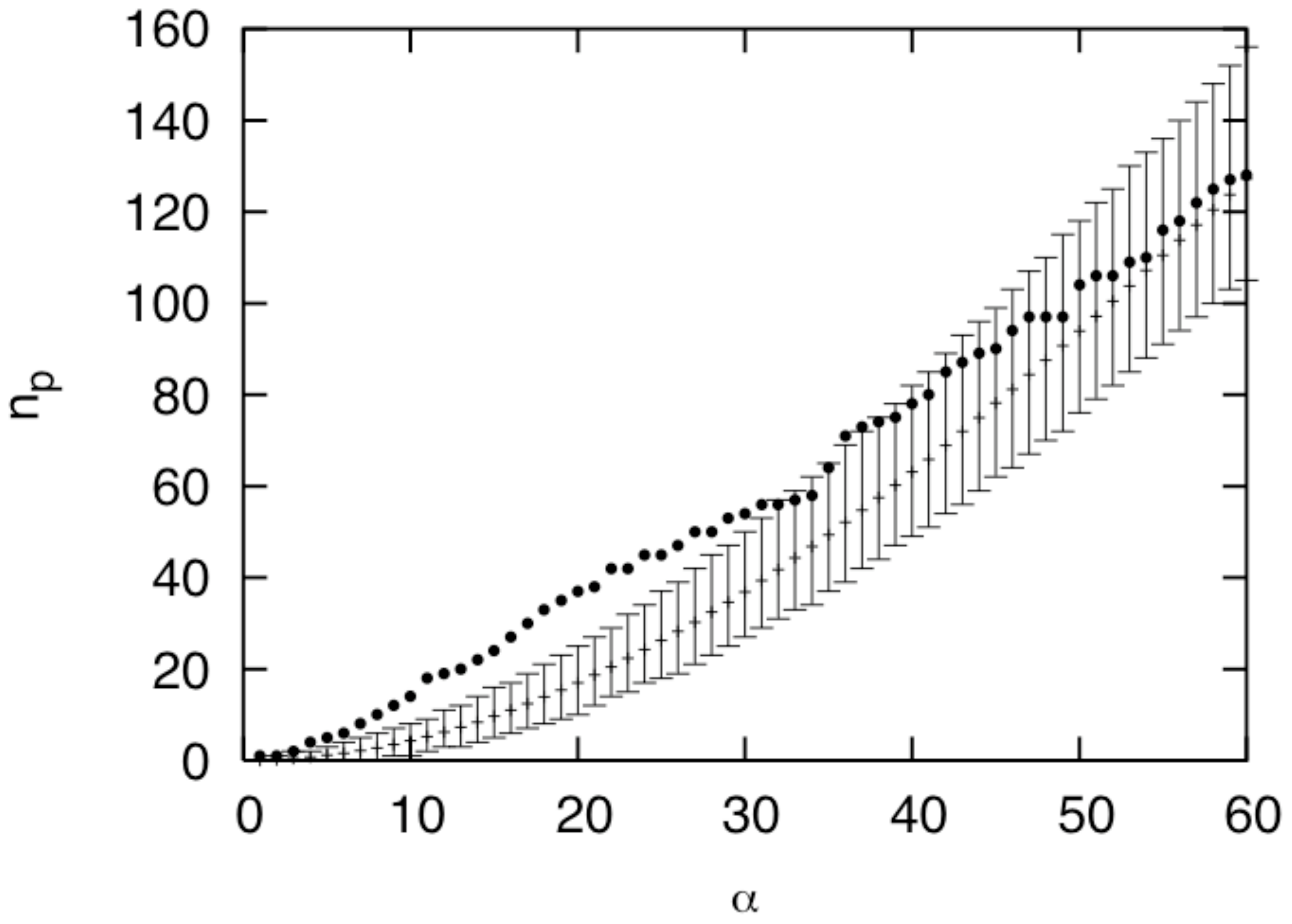}
\caption{Auto-correlation of the 27 highest energy events (dots) as a function of the 
angular scale in degrees. The expectations from isotropic samples of 27 directions is 
also shown (crosses) and the error bars represent the 90\% CL.}
\label{fig_acdi}
\end{figure}

Some idea of the anisotropic nature of the 27 most energetic events detected by 
the Auger Observatory might be obtained without using the V-C catalog, therefore avoiding 
the incompleteness issues related to this compilation of sources. One of the most simple 
intrinsic properties of a given set of arrival directions is given by its auto-correlation 
function, that is, the number of pairs for different angular scales. The auto-correlation 
for the highest events identified in the scan over the full data set is shown in figure 
\ref{fig_acdi}, together with the expectations from isotropic data sets. The departure 
from isotropy can be seen in a wide range of angles from 9$^{\circ}$ to 22$^{\circ}$, where 
the data shows a larger number of pairs than the isotropic expectations.

%

\section{Additional Remarks}

The topology of the galactic magnetic field is highly unknown. The few measurements of 
polarized light from galactic and extragalactic pulsars seems to point to a 
field of some $\mu$G with direction reversals between arms and an exponentially decreasing 
magnitude as a function of the distance to the galaxy center measured in the galactic 
plane \cite{Han:2006ci}. With the increasing statistics of the Auger surface detector, the prospects 
for the detection of multiplets are very promising, which could bring valuable information on 
the galactic field structure. However, a true knowledge of the field topology should be 
provided by a dedicated experiment, such as the Square Kilometer Array (SKA) \cite{SKApage}, 
which plans to measure about a million pulsar rotation curves, building a true 3-dimensional map 
of the regular magnetic field component in our galaxy. This kind of information, combined 
with the detection of multiplets, would enable us to measure particles rigidity dependence 
with energy and, in turn, get information on their chemical composition.

As already mentioned, the coincidence in energy between the flux attenuation and the 
enhancement of the anisotropy signal is an additional evidence for the GZK effect. Therefore, one 
could naively identify the maximum distance $D_{max}$ of sources coming from the scan with 
the horizon at $E_{min}$ (defined as the distance from us containing the sources responsible for 
90\% of the protons above a certain energy threshold.). At a first sight, then, the value of 
75 Mpc for $D_{max}$ as compared to $\sim 200$ Mpc \cite{Harari:2006uy} for the proton horizon 
at 60 EeV (assuming continuous energy losses, a uniform distribution of sources with 
equal intrinsec luminosities) could be seen as an inconsistency. Nonetheless, 
$D_{max}$ and the horizon are not even linearly correlated, chance correlations with 
foreground sources induce some bias on $D_{max}$ towards smaller source distances, and the 
same effect can be obtained in the presence of a large local overdensity. 

The HiRes Collaboration \cite{Abbasi:2008md} has performed a similar analysis, but it does not 
seem to confirm the Auger result. Above 56 EeV, from a total of 13 events, only 2 are correlated 
with AGNs with $z\le 0.018$. A difference in energy scale larger than the one envisaged by the 
Hires Collaboration could explain the discrepance if one believes the Northern and Southern skies 
are not so much different. On the other hand, an apparent lack of events in the region corresponding 
to the Virgo cluster 
seems to be a property of both data sets, an effect which could arise in a scenario where the bulk of the 
flux reaching us is generated in a very limited number of sources, like has been suggested in 
\cite{Gorbunov:2008ef}, in particular, analyzing the possibility of being Cen A the sole source. The 
construction of the Auger North \cite{Nitz:2007ur} will allow us to crosscheck the anisotropy results 
obtained from the South Observatory and from HiRes, to enhance the statistics in the highest energy 
part of the spectrum, look for multiplets, and even measure individual source spectra.

A detailed study of the sources correlating with the Auger high energy events, shows that 
they do not constitute the sub-sample with the most powerful sources. It is worth stressing 
though, that we do not claim the identification of the cosmic rays sources, and given the size of 
the correlation scale of less than $\sim 6^{\circ}$, it is impossible to pinpoint the true ones, 
and essentially, anything whose spatial distribution, when convoluted with the magnetic deflections, 
follows the AGNs clustering properties, cannot be excluded so far. 
An interesting possibility raised in \cite{Kotera:2008ae} is that, perhaps, some of the Auger 
events are pointing back to the last scattering centers, instead of their original sources, 
due to the presence of an inhomogeneously magnetized intergalactic medium.

There have been some concerns with respect to the incompleteness of the V-C catalog, close to 
the galactic plane (due to dust extinction), at high redshifts and the lack of a selection 
function, since this is in fact a compilation of surveys and not a survey by itself. However, the 
demonstration of the anisotropy is not affected by these problems, since the catalog 
has been used only as an intermediate step in order to establish the typical scale of the 
anisotropy signal.

\section{Conclusion}

The Pierre Auger Collaboration has established with more than 99\% CL the anisotropy 
of its highest energy sample. The reconstructed directions of showers with energy above 
60 EeV show a correlation on angular scales of less than $6^{\circ}$ with the positions 
of AGNs up to $\sim$100 Mpc from us. The evolution of the signal with the 
energy shows a sharp transition from 
isotropy to anisotropy at $E\sim 60$ EeV, and such a transition coincides with the flux 
attenuation observed in the Auger spectrum. This is an additional evidence for the 
existence of the GZK cutoff, where the abrupt reduction in the particles mean free 
path as the threshold for pion photo-production in the CMB is achieved, makes the 
universe essentially opaque at scales much larger than 100 Mpc for particles above 60 
EeV. As a byproduct, the small effective horizon at high energies enhances the 
anisotropies signal, revealing the clumpy aspects of the structures in our local universe.
Even though the correlation with the V-C catalog seems to be quite robust, the angular 
scale of $\sim 6^{\circ}$ does not make possible to unambiguously identify the sources. 
Essentially, anything which clusters (after taking into account magnetic deflections) in a 
similar way as AGNs are cannot be excluded as the true sources. Intrinsic (catalog independent) 
properties of the events, such as their auto-correlation function, show a clear departure from 
isotropy in a large angular range.





%

\end{document}